
\magnification
\magstep 1
\nopagenumbers
\vsize=22 true cm
\hsize=15 true cm
\parindent=1.2 true cm
\def\sa{\vskip 0.5 true cm}
\def\sb{\vskip 1 true cm}
\def\spar{\vskip 0.4 true cm}

\baselineskip=0.6 true cm

\rightline{\bf LYCEN 8766}

\sa

\rightline{November 1987}

\sb
\sb

\centerline{\bf Application of non-bijective transformations to
various potentials}

\sb
\sb
\sb

\centerline{\bf Maurice KIBLER}

\sb
\sa

\centerline{Institut de Physique Nucl\'eaire (et IN2P3)}

\centerline{Universit\'e Claude Bernard Lyon-1}

\centerline{43, Bd du 11 Novembre 1918}

\centerline{69622 Villeurbanne Cedex}

\centerline{France}

\sb
\sb

\noindent Paper presented at the XVIth International Colloquium on Group
Theoretical Methods in Physics (Varna, Bulgaria, 15-20 June 1987)
and published in ``Group Theoretical Methods in Physics'', edited by
H.-D. Doebner, J.-D. Hennig and T.D. Palev (Springer-Verlag, Berlin,
1988) : Lecture Notes in Physics {\bf 313}, 238 (1988).

\vfill\eject

\baselineskip=0.43 true cm

\centerline{\bf Application of non-bijective
transformations to various potentials}

\sb
\sb

\centerline{M. KIBLER}

\vskip 0.18 true cm

\centerline{Institut de Physique Nucl\'eaire (et IN2P3)}

\centerline{Universit\'e Claude Bernard Lyon-1}

\centerline{69622 Villeurbanne Cedex}

\centerline{France}

\sb
\sb

\centerline{ABSTRACT}

\vskip 0.38 true cm
\baselineskip = 0.45 true cm

Some results about non-bijective quadratic transformations generalizing the
Kustaanheimo-Stiefel and the Levi-Civita transformations are reviewed in
\S 1. The three remaining sections are devoted to new results:
\S 2 deals with the Lie algebras under constraints associated to
some Hurwitz transformations; \S 3 and \S 4 are concerned with several
applications of some Hurwitz transformations to wave equations for
various potentials in $R^3$ and $R^5$.

\baselineskip=0.588 true cm

\spar

\noindent {\bf 1. Non-bijective canonical transformations}

\vskip 0.23 true cm

We start with a $2m$-dimensional $(2m = 2, 4, 8, \ldots)$ Cayley-Dickson
algebra $A(c)$ where $c$ stands for a $p$-uple $(c_1, c_2, \ldots, c_p)$ such
that $c_i = \pm 1$ for $i = 1, 2, \ldots, p$ and $2^p = 2m$. (Remember that
$A(-1)$, $A(-1,-1)$ and $A(-1,-1,-1)$ are nothing but the algebras of complex
numbers, usual quaternions and usual octonions, respectively.) Let
$u = u_0 + \sum_{i=1}^{2m-1} u_i \ e_i$
be an element of $A(c)$ where $u_0, u_1, \ldots, u_{2m-1}$ are the components
of $u$ and $\{ e_1, e_2, \ldots, e_{2m-1} \}$ is a system of generators for
$A(c)$. We associate to the hypercomplex number $u$ the element $\hat u$ of
$A(c)$ defined by
$\hat u = u_0 + \sum_{i=1}^{2m-1} \epsilon_i \ u_i \ e_i$
where $\epsilon_i = \pm 1$ for $i = 1, 2, \ldots, 2m-1$. Let us consider the
right (or left) application
$A(c) \rightarrow A(c) \ : \ u \mapsto x = u \hat u \ ({\rm or} \ \hat u u)$.
Three cases can occur according to the form taken by $\hat u$.

\vskip 0.23 true cm

A. For $\hat u = u$: The (right = left) application $u \mapsto x = u^2$ defines
a map $R^{2m} \rightarrow R^{2m}$ which constitutes an extension of the
Levi-Civita map [1] corresponding to
$2m = 2$ and $c_1 = -1$ and of the map introduced in [2] and
corresponding to $2m = 4$ and $c_1 = c_2 = -1$. The maps $R^{2m} \rightarrow
R^{2m}$ for $2m = 2, 4, 8, \ldots$ correspond to the {\sl quasiHurwitz}
transformations of [3]. From a geometrical viewpoint, the map
$R^{2m} \rightarrow R^{2m}$ for fixed $2m$ is associated to a fibration on
spheres with discrete fiber in the compact case where
$c_i = -1$ $(i = 1, 2, \dots, p)$ and to a fibration on hyperboloids with
discrete fiber in the remaining non-compact cases.

\vskip 0.23 true cm

B. For $\hat u = j(u)$: By $j(u)$ we mean that the coefficients
$\epsilon_i$ $(i = 1, 2, \ldots, 2m-1$ are such that the application
$j \ : \ A(c) \rightarrow A(c) \ : \ u \mapsto \hat u = j(u)$ defines an
anti-involution of the algebra A(c). Then, the right (or left) application
$u \mapsto x = uj(u)$ (or $j(u)u$) defines a map
$R^{2m} \rightarrow R^{2m-n}$ with $n$
$(n = m -1 + \delta (m,1)$ or $2m - 1)$ being the number of zero components of
$x$. The latter map constitutes an extension of the Kustaanheimo-Stiefel map
[4] which corresponds to $2m = 4$, $n = 1$, $c_1 = c_2 = -1$ and, for example,
$\epsilon_1 = -\epsilon_2 = -\epsilon_3 = -1$. The map introduced by
Iwai [5] is obtained when $2m = 4$, $n = 1$, $c_1 = -c_2 = -1$ and, for
example, $\epsilon_1 = -\epsilon_2 = -\epsilon_3 = -1$. The maps
$R^{2m} \rightarrow R^{2m-n}$ for $2m = 2, 4, 8, \ldots$ correspond to the
{\sl Hurwitz} transformations of [3]. From a geometrical viewpoint, the
maps $R^{2m} \rightarrow R^{2m-n}$ for $2m = 2, 4, 8$ and $16$ are associated
to the classical Hopf fibrations on spheres with compact fiber in the compact
cases and to fibrations on hyperboloids with either
compact or non-compact fiber in the non-compact cases.

\vskip 0.23 true cm

C. For $\hat u \ne u$ or $j(u)$: This case does not lead to new
transformations for $2m =2$ and $4$. Some new transformations, referred to as
{\sl pseudoHurwitz} transformations in [3], arise for $2m \geq 8$. In
particular for $2m = 8$ and $\sum_{i=1}^{7} \epsilon_i = -3$ or $5$, the
application $u \mapsto u \hat u$ (or $\hat u u$) defines a map
$R^8 \rightarrow R^7$. Such a map is
associated to a Hopf fibration on spheres with compact fiber
for $(c_1,c_2,c_3) = (-1,-1,-1)$ and to fibrations on hyperboloids
with either compact or non-compact fiber for $(c_1,c_2,c_3) \ne (-1,-1,-1)$.

\vskip 0.23 true cm

The various transformations mentioned above may be presented in matrix
form. (A detailed presentation can be found in [3].)
Let us define the $2m \times 2m$ matrix
${\bf \epsilon} = diag(1, \epsilon_1, \epsilon_2, \ldots,
\epsilon_{2m-1})$
and let ${\bf u}$ and ${\bf x}$ be the $2m \times 1$ column-vectors
whose entries are the
components of the hypercomplex numbers $u$ and $x$, respectively. Then, the
$R^{2m} \rightarrow R^{2m-p}$ transformation defined by
$u \mapsto x = u \hat u$ (where $p = 0$ and
$2m = 2, 4, 8, \ldots$ for quasiHurwitz transformations, $p = m - 1 +
\delta(m,1)$ or $2m-1$ and $2m = 2, 4, 8, \ldots$ for Hurwitz transformations,
and $p \geq 1$ and $2m \geq 8$ for pseudoHurwitz transformations) can be
described by ${\bf x} = A({\bf u}) \ {\bf \epsilon} \ {\bf u}$ where
$A({\bf u})$ is a $2m \times 2m$ matrix. For $2m = 2, 4$ or $8$, the matrix
$A({\bf u})$ may be written in terms of Clifford matrices and constitutes an
extension of the Hurwitz matrices occurring in the Hurwitz factorization
theorem (see [3]). Note that there are $p$ zero entries in the
$2m \times 1$ column-vector ${\bf x}$ with, in particular, $p = 2m - 1$ or
$m - 1 + \delta(m,1)$ for Hurwitz transformations.

\spar

\noindent {\bf 2. Lie algebras under constraints}

\vskip 0.23 true cm

In this section, we shall restrict ourselves to the Hurwitz transformations
for $2m = 2, 4$ and $8$ and, more specifically, to the
$R^{2m} \rightarrow R^{2m-n}$ transformations
with $n = m - 1 + \delta (m,1)$. These
transformations are clearly non-bijective and this fact may be transcribed as
follows. Let us consider the $2m \times 1$ column-vector
$2 \ A({\bf u}) \ {\bf \epsilon} \ d{\bf u}$
where $d{\bf u}$ is the differential of the
$2m \times 1$ column-vector ${\bf u}$. It can be seen that
$2m-n$ components of
$2 \ A({\bf u}) \ {\bf \epsilon} \ d{\bf u}$
may be integrated to give the $2m-n$ non-zero
components of ${\bf x} = A({\bf u}) \ {\bf \epsilon} \ {\bf u}$. Further,
the remaining $n$ components of
$2 \ A({\bf u}) \ {\bf \epsilon} \ d{\bf u}$ are one-forms
$\omega_1, \omega_2, \ldots, \omega_n$ which are not total differentials. In
view of the non-bijective character of the map $R^{2m} \rightarrow R^{2m-n}$,
we can assume that $\omega_i = 0$ for $i = 1, 2, \ldots, n$. To each one-form
$\omega_i$, we may associate a vector field $X_i$ which is a bilinear form in
the $u_{\alpha}$ and $p_{\alpha} = \partial/\partial_{\alpha}$ for
$\alpha = 0, 1, \ldots, 2m-1$. An important property, for what follows, of
the latter vector fields is that $X_i\psi =0$ $(i = 1, 2, \ldots, n)$ for any
function $\psi$ of class $C^1(R^{2m-n})$ and $C^1(R^{2m})$ in the variables
of type $x$ and $u$, respectively. In addition, it can be verified that the
$n$ operators $X_i$ $(i = 1, 2, \ldots, n)$ span a Lie algebra. We denote
$L_0$ this algebra and refer it to as the {\sl constraint Lie algebra}
associated to the $R^{2m} \rightarrow R^{2m-n}$ Hurwitz transformation.

Now, it is well known that the $2m(4m + 1)$ bilinear forms
$u_{\alpha}u_{\beta}$, $u_{\alpha}p_{\beta}$ and $p_{\alpha}p_{\beta}$ for
$\alpha, \beta = 0, 1, \ldots, 2m-1$ span the real symplectic Lie algebra
$sp(4m,R)$ of rank $2m$. We may then ask the question: what remains of the
Lie algebra $sp(4m,R)$ when we introduce the $n$ constraint(s) $X_i = 0$
$(i = 1, 2, \ldots, n)$ into $sp(4m,R)$. (It should be noted that each
constraint $X_i = 0$ may be regarded as a {\sl primary constraint}
in the sense of
Dirac, cf. [6,7].) This amounts in last analysis to look for the
centralizer of $L_0$ in $L = sp(4m,R)$ [8]. The resulting Lie algebra
$L_1 = cent_LL_0/L_0$ is referred to as the {\sl Lie algebra under
constraints} associated to the $R^{2m} \rightarrow R^{2m-n}$ Hurwitz
transformation. An important result is the following.

{\sl Result 1}. For fixed $2m$, the constraint Lie algebra $L_0$ and the Lie
algebra under constraints $L_1$ are characterized by the (compact or
non-compact) nature of the fiber of the fibration associated to the
$R^{2m} \rightarrow R^{2m-n}$ Hurwitz transformation.

The determination of $L_1$ has been achieved, in partial form,
for one of the cases $(2m,2m-n) = (4,3)$ in [9] and, in complete form,
for all the cases $(2m,2m-n) = (2,1), (4,3)$ and $(8,5)$ in [8].

\spar

\noindent {\bf 3. Generalized Coulomb potentials in $R^3$ and $R^5$}

\vskip 0.23 true cm

Let us begin with the ``Coulomb'' potential ($-Ze^2$ is a coupling constant):
$$V_5 = -Ze^2/(x_0^2 - c_2 x_2^2 + c_1 c_2 x_3^2 - c_3 x_4^2
                                 + c_1 c_3 x_5^2)^{1/2}$$
in $R^5$ equipped with the metric $\eta_5 = diag(1,-c_2,c_1 c_2,-c_3,c_1 c_3)$
and consider the Schr\"odinger equation for this potential and this metric.
(The corresponding generalized Laplace operator is
$\tilde {\bf \nabla} \eta_5 {\bf \nabla}$ in the variables
$(x_0, x_2, x_3, x_4, x_5)$.) We now
apply a $R^8 \rightarrow R^5$ Hurwitz transformation to the
considered problem. The knowledge of the transformation properties of
the generalized Laplace operators under the $R^8 \rightarrow R^5$ Hurwitz
transformations leads to the following result.

{\sl Result 2}. The Schr\"odinger equation for the potential $V_5$ in $R^5$
with the metric $\eta_5$
and the energy $E$
is equivalent to a set consisting of (i) one
Schr\"odinger equation for the harmonic oscillator potential
$$V_8 = -4E(u_0^2 - c_1 u_1^2 - c_2 u_2^2 + c_1 c_2 u_3^2 - c_3 u_4^2
                    + c_1 c_3 u_5^2 + c_2 c_3 u_6^2 - c_1 c_2 c_3 u_7^2)$$
in $R^8$ with the metric
$\eta_8 = diag(1,-c_1,-c_2,c_1c_2,-c_3,c_1c_3,c_2c_3,-c_1c_2c_3)$
and the energy $4Ze^2$
and (ii) three first-order differential equations associated to the $n = 3$
constraints of the $R^8 \rightarrow R^5$ Hurwitz transformations.

A similar result is obtained under the evident replacements:
$V_5 \rightarrow V_3 = -Ze^2/(x_0^2 - c_2 x_2^2 + c_1 c_2 x_3^2)^{1/2}$,
$\eta_5 \rightarrow \eta_3 = diag(1,-c_2,c_1 c_2)$,
$V_8 \rightarrow V_4 = -4E(u_0^2 - c_1 u_1^2 - c_2 u_2^2 + c_1 c_2 u_3^2)$,
$\eta_8 \rightarrow \eta_4 = diag(1,-c_1,-c_2,c_1 c_2)$ and $n = 3
\rightarrow n = 1$.
The so-obtained result for the generalized Coulomb potential $V_3$ in $R_3$
equipped with the metric $\eta_3$ thus corresponds to $c_3 = 0$. Note that
the usual Coulomb potential $-Ze^2/(x_0^2 + x_2^2 + x_3^2)^{1/2}$
corresponds to $c_3 = 0$ and $c_1 = c_2 = -1$. (The case of $V_3$
with $c_1 = c_2 = -1$ has been investigated
in [10,11,12] and the case of $V_5$ with $c_1 = c_2 = c_3 = -1$
has been recently considered in [3,12,13].)

As a corollary, information on the spectrum of the hydrogen atom in $R^5$
($R^3$) with the metric $\eta_5$ ($\eta_3$) can be deduced from the knowledge
of the spectrum of the harmonic oscillator in $R^8$ ($R^4$) with the metric
$\eta_8$ ($\eta_4$). By way of illustration, we shall continue with
$c_1 = c_2 = c_3 = -1$ (i.e., the case of a Coulomb potential in $R^5$ with
unit metric), on the one hand, and with $c_1 = c_2 =  c_3 -1 = -1$ (i.e., the
case of the usual Coulomb potential), on the other hand. The non-invariance
dynamical algebras for the corresponding isotropic harmonic oscillators in
$R^8$ and $R^4$ are clearly $sp(16,R)$ and $sp(8,R)$, respectively. Then, the
non-invariance dynamical algebras for the corresponding hydrogen
atoms in $R^5$ and $R^3$ are nothing but the Lie algebras under constraints
$L_1$ associated to the $R^8 \rightarrow R^5$ and $R^4 \rightarrow R^3$
compact
Hurwitz transformations, respectively. The results for $L_1$ of [8,9]
yield the non-invariance dynamical algebras $L_1 = so(6,2)$ and $so(4,2)$ for
the hydrogen atoms in $R^5$ and $R^3$, respectively.

To close this section, let us show how to obtain the discrete spectra for the
$R^5$ and $R^3$ hydrogen atoms under consideration. A careful examination of
the hydrogen-oscillator connection shows that energies and coupling constants
are exchanged in such a connection. As a matter of fact, we have
$$(1/2) \mu (2 \pi \nu)^2 = - 4E \qquad
h \nu (\sum_{\alpha = 0}^{2m-1} n_{\alpha} + m) = 4 Ze^2 \qquad
n_{\alpha} \in N \qquad
2m = 8 \ {\rm or} \ 4$$
where $\mu (2 \pi \nu)^2$ is
the coupling constant for the oscillator (whose mass $\mu$ is the
reduced mass of the Coulomb system)
and $n_{\alpha}$ for $\alpha = 0, 1, \ldots, 2m-1$ are the (Cartesian)
quantum numbers for the isotropic harmonic oscillator in $R^{2m}$ ($2m = 8$ or
$4$). Furthermore, the $n$ constraint(s) associated to the $R^{2m}
\rightarrow R^{2m-n}$ Hurwitz transformations for $2m = 8$ and $4$ yield
$\sum_{\alpha = 0}^{2m-1} n_{\alpha} + 2 = 2k$
where $k$ ($= 1, 2, 3, \ldots$) plays the role of a principal quantum number
(cf. [11]). By eliminating
the frequency $\nu$ from the formulas connecting coupling constants and
energies, we end up with
$$E = E_0/(k + m/2 - 1)^2 \qquad E_0 = - \mu Z^2 e^4/(2 \hbar^2) \qquad
k = 1, 2, 3, \ldots$$
where $m/2 = 1$ and $2$ for the $R^3$ and $R^5$ hydrogen atoms, respectively,
in agreement with the Bohr-Balmer formula in arbitrary dimension (see, for
example, [7,14]).

\spar

\noindent {\bf 4. Axial potentials in $R^3$}

\vskip 0.23 true cm

A. Generalized Hartmann potential in $R^3$. Let us
consider the potential
$$W_3 = -\eta \sigma^2 / r + (1/2) q \eta^2 \sigma^2 / \rho^2$$
in $R^3$ equipped with the metric $\eta_3$. The variables $r$ and $\rho$ are
``distances'' in $R^3$ and $R^2$ given by
$r = (x_0^2 - c_2 x_2^2 + c_1 c_2 x_3^2)^{1/2}$ and
$\rho = (-c_2 x_2^2 + c_1 c_2 x_3^2)^{1/2}$, respectively.
The parameters $\eta$ and $\sigma$ are positive and the parameter $q$ is
such that $0 \leq q \leq 1$. The potential $W_3$ is an extension of the
so-called Hartmann potential which is of interest in the quantum chemistry of
ring-shaped molecules. Indeed, the
Hartmann potential corresponds to $c_1 = c_2 = -1$ and $q = 1$ (cf.
[15]). The potential $W_3$ for $q = 1$ and $(c_1,c_2)$ arbitrary shall be
called generalized Hartmann potential. It is to be observed that for
$q = 0$ and $\eta \sigma^2 = Z e^2$, the potential $W_3$ identifies to the
(generalized) Coulomb potential $V_3$ in $R^3$ equipped with the metric
$\eta_3$. (The parameter $q$ is thus simply a distinguishing parameter which,
for $0 \leq q \leq 1$, may be restricted to take the values $0$ or $1$.)

It is possible to find an $R^4 \rightarrow R^3$ Hurwitz transformation to
transform the $R^3$
Schr\"odinger equation, with the metric $\eta_3$,
for the generalized Hartmann potential into an $R^4$
Schr\"odinger equation, with the metric $\eta_4$,
for a non-harmonic oscillator
plus a constraint equation. Each of the two obtained equations can be separated
into two $R^2$ equations. This leads to the following result where $E$ denotes
the energy of a particle of (reduced) mass $\mu$ in the potential $W_3$. Such
a result generalizes the one derived in [15] for the special case $c_1 = c_2
= -1$.

{\sl Result 3}. The $R^3$ Schr\"odinger equation, with the metric $\eta_3$,
for the generalized Hartmann potential $W_3$ is equivalent to a set comprizing
(i) two coupled $R^2$ Schr\"odinger equations for two two-dimensional
oscillators with mass $\mu$, one with the metric $diag(1,-c_1)$ and the
potential
$V_{01}=-4E(u_0^2 - c_1 u_1^2)
+ (1/2) q \eta^2 \sigma^2 / (u_0^2 - c_1 u_1^2)$,
the other with the metric $diag(-c_2,c_1 c_2)$ and the potential
$V_{23}=-4E(-c_2u_2^2+c_1c_2u_3^2)
+ (1/2)q\eta^2\sigma^2/(-c_2u_2^2+c_1c_2u_3^2)$,
and (ii) two coupled $R^2$ constraint equations.

\vskip 0.23 true cm

B. Generalized Coulomb + Aharonov-Bohm potential in $R^3$. Let us consider
$$X_3 = Ze'e''/r + (2 \mu \rho^2)^{-1}[A + iB(x_2 \partial / \partial x_3
+ c_1 x_3 \partial / \partial x_2)]$$
in $R^3$ equipped with the metric $\eta_3$. Here again, we have
$r = (x_0^2 - c_2 x_2^2 + c_1 c_2 x_3^2)^{1/2}$ and
$\rho = (-c_2 x_2^2 + c_1 c_2 x_3^2)^{1/2}$. The generalized Hartmann
potential $W_3$ can be obtained as a special case of $X_3$: when
$Ze'e'' = - \eta \sigma^2$, $A/ \mu = q \eta^2 \sigma^2$ and $B = 0$, the
potential (energy)
$X_3$ identifies to $W_3$. In the (compact) case $c_1 = c_2 = -1$
and for $A = (e'f/c)^2$ and $B = 2 e' \hbar f/c$ with $f = F/(2 \pi)$, the
(velocity-dependent) operator
$X_3$ describes the interaction of a particle of charge $e'$ and (reduced)
mass $\mu$ with a potential $({\bf A},V)$, where the scalar potential
$V = Z e''/(x_0^2 + x_2^2 + x_3^2)^{1/2}$ is of the Coulomb type and the
vector potential
${\bf A} = (A_{x_2} = -[x_3 / (x_2^2 + x_3^2)] f,
            A_{x_3} =  [x_2 / (x_2^2 + x_3^2)] f,
            A_{x_0} =  0)$
is of the Aharonov-Bohm type (cf. [16]). We are now in a position to
list the following result which generalizes the one obtained in [16] for
the special case $c_1 = c_2 = -1$.

{\sl Result 4}. It is possible to find a Hurwitz transformation to convert the
$R^3$ Schr\"odinger equation, with the metric $\eta_3$,
for $X_3$ into an $R^4$ Schr\"odinger equation, with the metric $\eta_4$,
accompanied by a constraint condition. The separation of variables from
$R^4$ to $R^2 \times R^2$ is possible here again and this leads to a result
similar to Result 3 with the replacements
$V_{01} \rightarrow
-4E(u_0^2 - c_1 u_1^2) + (A - mB)/[2 \mu (u_0^2 - c_1 u_1^2)]$ and
$V_{23} \rightarrow
-4E(-c_2 u_2^2+c_1 c_2 u_3^2)+(A-mB)/[2\mu(-c_2 u_2^2+c_1 c_2 u_3^2)]$,
with $im$ being a separation constant.

\vskip 0.23 true cm

C. Coulomb + Sommerfeld + Aharonov-Bohm + Dirac potential in $R^3$. We close
this paper with a brief study of the potential (energy):
$$Y_3 = Ze'e''/r + s/r^2 + (2 \mu \rho^2)^{-1}[\alpha ({\bf x}) +
i \beta ({\bf x})(x_2 \partial / \partial x_3 - x_3 \partial / \partial x_2)]$$
in $R^3$ equipped with the unit metric. The
distances $r$ and $\rho$ are given here by
$r = (x_0^2 + x_2^2 + x_3^2)^{1/2}$ and $\rho = (x_2^2 + x_3^2)^{1/2}$. The
potential $Y_3$ includes a Coulomb term $Ze'e''/r$ and a Sommerfeld term
$s/r^2$. Further, we take $\alpha ({\bf x}) = [(e'/c) f({\bf x})]^2$ and
        $\beta  ({\bf x}) = (2 e' \hbar / c) f({\bf x})$ with
                                 $f({\bf x}) = F/(2 \pi) + g(1 - x_0/r)$ so
that the two other terms in $Y_3$ describe the interaction of a particle
of charge $e'$ and (reduced) mass $\mu$ with the vector potential
${\bf A} = (-[x_3 / (x_2^2 + x_3^2)] f({\bf x}),
             [x_2 / (x_2^2 + x_3^2)] f({\bf x}),
             0)$
where $F$ refers to an Aharonov-Bohm potential and $g$ to a Dirac monopole
potential. (The vector potential ${\bf A}$
corresponds to the magnetic field
${\bf B} = g {\bf r} / r^3$). It is to be noted that the potential $Y_3$ with
$s = g = 0$ yields the potential $X_3$ with $c_1 = c_2 = -1$.

We can apply a compact Hurwitz transformation to the $R^3$ Schr\"odinger
equation for the potential $Y_3$. This leads to a system comprizing an $R^4$
Schr\"odinger equation and a constraint equation. The latter system is not
always separable from $R^4$ to $R^2 \times R^2$. Separability is obtained for
$2 \mu s = (e'g/c)^2$. This may be precised with the following preliminary
result to be developed elsewhere.

{\sl Result 5}. The $R^3$ Schr\"odinger equation, with the
usual metric $diag(1,1,1)$,
for the potential $Y_3$ with $2 \mu s = (e'g/c)^2$ is equivalent to a set
comprizing (i) two coupled $R^2$ Schr\"odinger equations for two
two-dimensional isotropic oscillators involving each a centrifugal term and
(ii) two coupled $R^2$ constraint equations.

\spar

\noindent {\bf References}

\vskip 0.23 true cm
\parindent = 0.9 true cm
\baselineskip = 0.5 true cm

\item{[1]} Levi-Civita, T., Opere Matematiche {\bf 2} (1906); Acta. Math.
{\bf 42}, 99 (1920).

\item{[2]} Kibler, M. and N\'egadi, T., Croat. Chem.
Acta, CCACAA, {\bf 57}, 1509 (1984).

\item{[3]} Lambert, D. and Kibler, M., Preprint Lycen 8642 (IPN de Lyon, 1986).

\item{[4]} Kustaanheimo, P. and Stiefel, E.,
J. reine angew. Math. {\bf 218}, 204 (1965).

\item{[5]} Iwai, T., J. Math. Phys. {\bf 26}, 885 (1985).

\item{[6]} Dirac, P.A.M., {\sl Lectures on Quantum Mechanics} (Yeshiva
University: New York, 1964).

\item{[7]} Todorov, I.T., Ann. Inst. Henri Poincar\'e, Sect. A,
{\bf 28}, 207 (1978).

\item{[8]} Kibler, M. and Winternitz, P.,
Preprint Lycen 8755 (IPN de Lyon, 1987).

\item{[9]} Kibler, M. and N\'egadi, T., Lett. Nuovo Cimento {\bf 37}, 225
(1983); J. Phys. A: Math. Gen. {\bf 16},
4265 (1983); Phys. Rev. A {\bf 29}, 2891 (1984).

\item{[10]} Boiteux, M., C. R. Acad. Sci. (Paris), Ser. B,
{\bf 274}, 867 (1972).

\item{[11]} Kibler, M. and N\'egadi, T., Lett. Nuovo Cimento {\bf 39},
319 (1984); Theoret. Chim. Acta {\bf 66}, 31 (1984). In these papers, the
(uncorrect) relation $n_1 + n_2 = n_3 + n_4$ should be replaced by
$n_1 + n_2 + n_3 + n_4 + 2 = 2k$ ($k = 1, 2, 3, \ldots$).

\item{[12]} Kibler, M., Ronveaux, A. and N\'egadi, T., J. Math. Phys.
{\bf 27}, 1541 (1986).

\item{[13]} Davtyan, L.S., Mardoyan, L.G., Pogossyan, G.S., Sissakyan,
A.N. and Ter-Antonyan, V.M., Preprint P5-87-211 (JINR: Dubna, 1987).

\item{[14]} Mladenov, I.M. and Tsanov, V.V., C. R. Acad. bulgare Sci. (Sofia)
{\bf 39}, 35 (1986); J. Geom. and Phys. {\bf 2}, 17 (1985).

\item{[15]} Kibler, M. and N\'egadi, T., Int. J. Quantum Chem. {\bf 26},
405 (1984); Kibler, M. and
Winternitz, P., J. Phys. A: Math. Gen. {\bf 20}, 4097 (1987).

\item{[16]} Kibler, M. and N\'egadi, T., Phys. Lett. A {\bf 124}, 42
(1987).

\bye